\documentclass[aps,pra,twocolumn,superscriptaddress,nofootinbib,10pt]{revtex4-2}
\usepackage{graphicx}
\usepackage{enumitem}
\usepackage{amssymb,amsfonts,amsthm,mathtools,mathrsfs}
\usepackage{bm}
\usepackage{xcolor}
\usepackage{subfig}
\usepackage{qcircuit}
\usepackage{booktabs}
\usepackage{multirow}
\usepackage{makecell}
\usepackage{ragged2e}
\usepackage[colorlinks=true,citecolor=blue,linkcolor=blue,urlcolor=blue]{hyperref}
\makeatother
\newcommand{\subfig}[2]{%
    {\footnotesize\textsf{\textbf{#1}}} 
    \vtop{
    \kern 5pt
    \hbox{\hskip -20pt #2}
}}

\begin{document}
\title{Optimizing Atom Transport, Gate-Count and Depth with Parity Twine}

\author{Javad Kazemi}
\affiliation{Parity Quantum Computing Germany GmbH, 20095 Hamburg, Germany}
\author{Michael Fellner}
\affiliation{Parity Quantum Computing GmbH, A-6020 Innsbruck, Austria}
\author{Riccardo J. Valencia-Tortora}
\affiliation{Parity Quantum Computing GmbH, A-6020 Innsbruck, Austria}
\author{Michael Schuler}
\affiliation{Parity Quantum Computing GmbH, A-6020 Innsbruck, Austria}
\author{Wolfgang Lechner}
\affiliation{Parity Quantum Computing Germany GmbH, 20095 Hamburg, Germany}
\affiliation{Parity Quantum Computing GmbH, A-6020 Innsbruck, Austria}
\affiliation{Institute for Theoretical Physics, University of Innsbruck, A-6020 Innsbruck, Austria}
\date{\today}

\begin{abstract}
We present an efficient implementation of the Parity Architecture for neutral-atom quantum processors. 
We adapt Parity Twine Networks (PTNs) to different atom layouts, native entangling gates, and atom-shuttling capabilities. This provides a general framework for hardware-aware optimization of gate count, circuit depth, and atom transport for quantum circuits encoding arbitrary interaction graphs in a common basis.
Specifically, we develop PTN constructions based on different native entangling-gate realizations, namely CZ, CZSWAP, and iSWAP, providing flexibility to accommodate different hardware capabilities on both static and mobile neutral-atom platforms.
Using the quantum Fourier transform (QFT) as a representative example, we demonstrate substantial reductions in two-qubit gate count, atom transport, and circuit depth. 
These resource savings translate into an estimated circuit fidelity three orders of magnitude higher than competing compilation strategies for a 30-qubit QFT.
We further extend the construction to the recently introduced optimistic QFT and discuss the broader applicability of PTNs to other quantum algorithms on neutral-atom platforms.

\end{abstract}

\maketitle

\section{Introduction}
Neutral-atom quantum computing~\cite{Saffman2010, Browaeys2020, Henriet2020, Menssen2026} is a promising platform for scalable quantum computation, combining high-fidelity Rydberg-mediated entangling gates~\cite{Levine2019, Jandura2022, Muniz2024, Radnaev2025, Giudici2025, Kazemi2025, Evered2026, Bergonzoni2026, Ildefonso2026} with flexible control over the spatial arrangement of qubits~\cite{Barredo2018, Song2021, Bluvstein2022, Norcia2023, Bluvstein2024, Rines2026, Bluvstein2026, Lib2026}. In particular, atoms can be arranged in different geometries and, in shuttling-capable architectures, transported during computation to dynamically modify their relative positions. Atom movement therefore provides more than a means of increasing effective connectivity: it can also enable motion-assisted and composite gate operations that are difficult or impossible to realize in a static architecture~\cite{Shaw2024, Ginzel2026, Lib2026}, facilitate the simulation of complex Hamiltonian systems~\cite{Kalinowski2023, Maskara2025, Maskara2025b, Evered2025}, and reduce resource overheads in fault-tolerant quantum computation~\cite{Bluvstein2022, Bluvstein2024, Reichardt2025, Rines2026, Bhardwaj2026, Bluvstein2026}. 

This additional degree of freedom comes at the cost of relatively slow transport operations and associated errors, introducing new compilation trade-offs involving entangling-gate count, circuit depth, and atom movement~\cite{Constantinides2024, Wang2024, Schmid2024, Ludmir2024, Tan2024, Tan2025, Huang2025, Gao2025}. 
The optimal use of atom movement therefore depends on how transport can be integrated with the native gate set and the structure of the target quantum circuit. This motivates compilation strategies that explicitly exploit both the connectivity and the mobility of the underlying hardware.

\begin{figure}[t!]
\includegraphics[width=0.95\columnwidth]{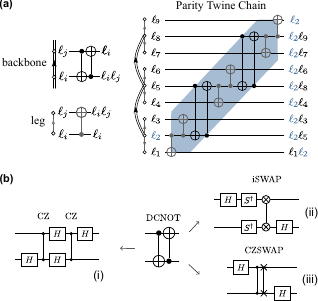}
\caption{(a) Information flow in a Parity Twine chain (highlighted in blue) with a backbone (black double lines) and two legs (grey lines) per backbone node. {DCNOT} gates are applied along the backbone, while CNOT gates connect the backbone to the legs, systematically distributing the logical parity information, labeled by $\ell_i$, throughout the system. (b) Equivalent decompositions of the DCNOT gate using neutral-atoms-native CZ, CZSWAP, or iSWAP gates. 
}
\label{fig:ptc}
\end{figure}

In this work, we build upon Parity Twine Networks (PTNs)~\cite{Dreier2025, Klaver2026}, a connectivity-aware framework for compiling quantum circuits with dense interaction structures, and extend their use to shuttling-capable neutral-atom architectures. PTNs originate from the Parity Architecture~\cite{Lechner2015, Fellner2022a} and exploit parity encoding to realize logical multi-qubit interactions through single-qubit rotations while distributing the required parity information using CNOT-based networks~\cite{Dreier2025}. Their structure can be embedded in different qubit layouts and adapted to different native entangling gates [cf. Fig.~\ref{fig:ptc}], making PTNs well suited to platforms with flexible connectivity and gate sets. In particular, we exploit atom movement to realize a composite CZSWAP operation, in which a Rydberg-mediated CZ gate is combined with a physical exchange of the participating atoms~\cite{Gao2025}. This allows physical exchange to replace entangling gates that would otherwise be required during compilation, turning atom transport from a routing overhead into a computational resource.

The efficiency of PTNs has recently been demonstrated across different hardware platforms, including trapped-ion architectures~\cite{Dominguez2024} and superconducting quantum computing platforms, where PTNs have been applied to the quantum Fourier transform (QFT)~\cite{Aumann2026}. In this work, we extend this approach to neutral-atom quantum computers (NAQCs). We develop PTN implementations tailored to the underlying hardware, accounting for static and mobile architectures, different qubit layouts, and native entangling gates. This hardware-aware co-design allows us to optimize both the quantum circuit and the physical operations required to execute it.
Our PTN-based QFT implementations achieve gate counts not previously demonstrated on neutral-atom platforms, yielding estimated circuit fidelities orders of magnitude higher than those of competing compilation approaches. We further extend the construction to the optimistic QFT (OQFT), whose block structure enables reduced circuit depth through parallelism and approximate commutativity~\cite{Kahanamokumeyer2025}.

These results demonstrate how hardware-aware compilation can exploit the interplay between connectivity, entangling-gate resources, circuit depth, and atom transport to substantially improve QFT performance on neutral-atom quantum computers. While the QFT provides a particularly clear example of a dense all-to-all interaction structure, the PTN framework has broader applicability and can also be applied to other algorithms and subroutines involving dense interactions, ranging from quantum arithmetic to variational algorithms such as QAOA.

The remainder of this paper is organized as follows. In Sec.~\ref{sec:PTN}, we review the core principles of PTNs and describe how this framework adapts to different qubit layouts and connectivity. In Sec.~\ref{sec:PTN_NAQC}, we briefly review neutral-atom quantum computing and present PTN-based hardware-aware compilation strategies for neutral-atom platforms. Sec.~\ref{sec:Applications} then presents a detailed evaluation of exact QFT implementations, benchmarking against competing compilation approaches, and an extension to the optimistic QFT. Finally, Sec.~\ref{sec:outlook} summarizes the main results and discusses future directions, including hardware-aware compiler optimizations and extensions of the PTN framework to other applications.

\section{Parity Twine Networks}
\label{sec:PTN}

PTNs are particularly well suited for applications that require an effectively all-to-all interaction structure~\cite{Dreier2025, Klaver2026, Aumann2026}. A key example is the QFT, a central subroutine in numerous quantum algorithms.
The Parity Twine framework is built from CNOT-based building blocks, referred to as \emph{Parity Twine chains}, or \textit{Twine chains} for short. To describe the logical information encoded by these CNOT networks, we assign to each physical qubit $j$ a logical parity label $\ell_j$. Throughout this work, we use $\ell_j$ to denote the logical $Z$-parity encoded on qubit $j$. In the $Z$ basis, a CNOT encodes the $Z$-parity information of its control and target qubits on the target qubit~\cite{Dreier2025, Klaver2026b}, as shown in Fig.~\ref{fig:ptc}(a).
Each Twine chain distributes quantum information and generates all two-body parity labels $\ell_i \ell_j,\ j>i$ associated with a single qubit $i$ [see Fig.~\ref{fig:ptc}(a) for an example]. A sequence of $n-1$ consecutively shorter Twine chains, with $n$ denoting the total number of qubits, collectively generates the complete set of two-body parity labels $\ell_i \ell_j, \forall i,j \in \{1, ..., n\}, j\neq i$. 
Applying a physical rotation $R_Z^{(k)}(\theta)$ on qubit $k$ carrying the parity label $\ell_i \ell_j$ effectively corresponds to performing the logical rotation $\bar{R}_{Z_i Z_j}(\theta) = e^{-i \theta Z_i Z_j / 2}$ on the qubits $i$ and $j$.
Thus, the resulting parity encoding enables performing all two-qubit interactions through single-qubit rotations~\cite{Klaver2026b} and, in particular, minimizes both gate count and circuit depth~\cite{Dreier2025}.

\begin{figure}[t!]
\includegraphics[width=\columnwidth]{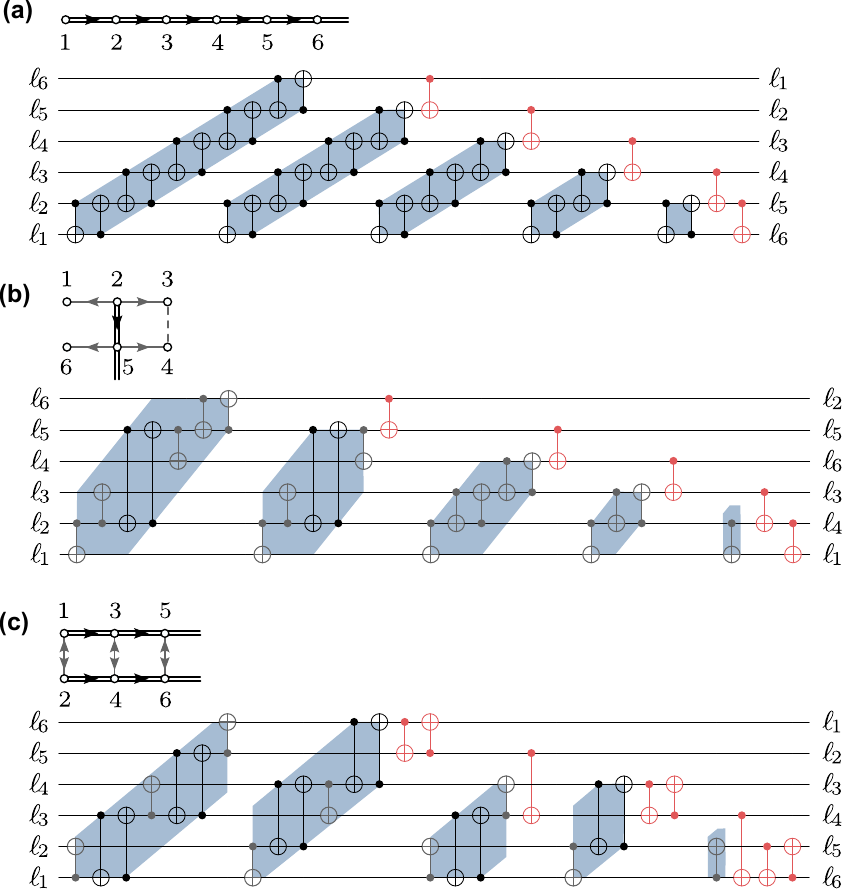}
\caption{Construction of Parity Twine networks as a cascade of Parity Twine chains (highlighted in blue) for (a) linear, (b) square, and (c) ladder layouts from a given Twine path (shown above each circuit). Black DCNOT gates implement the backbones, gray CNOT gates the legs, and red CNOT gates the decoding sequence that recovers the individual qubit labels at the end. Note that the circuits are stretched horizontally for visualization purposes to better highlight the structure of the Parity Twine networks and can be parallelized in practice. Therefore, they do not indicate their relative execution times.
}
\label{fig:ptn_circuits}
\end{figure}

A PTN can be embedded in qubit layouts with diverse connectivity and actively leverages increased connectivity to reduce entangling gate counts, making it a flexible tool for circuit synthesis. 
In the most restrictive setting, namely, a linear chain with nearest-neighbor interactions only, Twine chains reduce to a sequence of DCNOT gates, i.e., two consecutive CNOT gates applied in opposite directions [cf.~Fig.~\ref{fig:ptc}(b)]. 
Here, a PTN is given by a hierarchy of Twine chains of progressively decreasing size, followed by a decoding stage consisting of CNOT gates that recover logical single-qubit information at the end of the PTN circuit, as illustrated in Fig.~\ref{fig:ptn_circuits}(a).

This basic construction generalizes to higher-dimensional architectures, including quasi-one-dimensional ladder layouts, two-dimensional square lattices, and fully connected all-to-all layouts.
In such settings, two building blocks in a PTN are distinguished: \emph{backbones} and \emph{legs}.
A \emph{backbone} connection is implemented via DCNOT gates, while a \emph{leg} connection is implemented via single CNOT gates. The two can be combined to efficiently generate the full PTN. 
For the linear chain, discussed above, the PTN consists only of a single backbone going through the entire chain, without any legs. 
On the other hand, on a square lattice, for example, a PTN can be given by arranging the \emph{backbone} in a striped pattern, running along every other column or row of the lattice. The qubits in the intervening columns or rows are then connected to the backbone by  \emph{legs} [see Fig.~\ref{fig:ptc}(a) and Fig.~\ref{fig:ptn_circuits}(b)].

Another interesting layout for the PTN is the \emph{ladder} [cf.~Fig.~\ref{fig:ptn_circuits}(c)], in which two parallel backbones are connected through transverse legs. 
The ladder architecture is particularly notable in that it offers a favorable trade-off among the considered layouts, achieving the minimum circuit depth while requiring only a modest increase in CNOT gate count relative to the square layout ($3/4n^2$ versus $2/3n^2$ in leading order).
A practical limitation, however, is that the ladder is inherently a quasi-one-dimensional structure, and cannot be easily bent or folded into a 2D lattice. 
In neutral-atom architectures, however, the flexibility in arranging qubits in arbitrary geometries and transporting them provides native ways to realize the ladder in two dimensions. 
One approach would be to bend the ladder into a two-dimensional layout by carefully placing atoms at the boundaries in a non-regular arrangement.
Another approach would be to break the ladder into linear segments, implementing the missing links between segments with short-range shuttling operations. 
We exploit this flexibility in the construction of the optimistic QFT discussed in Sec.~\ref{sec:oqft}.

The inclusion of legs, which only require single CNOTs\footnote{In some cases, such as the square layout, restoring the parity labels on the legs may require DCNOTs rather than single CNOTs.}, compared to the DCNOTs on the backbone, immediately reduces the CNOT gate count.
However, this comes at the cost of modestly increased depth of the PTN compared to the linear PTN, because consecutive Twine chains need more delay before they can be started~\cite{Dreier2025}, as reflected by the increased width of the Twine chains in Fig.~\ref{fig:ptn_circuits}.
In general, increased connectivity reduces the overall gate count, but it does not necessarily imply a reduction in circuit depth~\cite{Dreier2025}.
This trade-off is common to PTNs with legs, making it nontrivial to determine a priori which layout is optimal: the best choice depends on the relative importance of gate count and circuit depth for the underlying hardware and application.

Having introduced the Parity Twine framework for a highly efficient compilation of problems with all-to-all interaction structure, in the following section we design tailored implementations of PTNs for NAQCs, focusing on monolithic platforms where all operations are performed in a single zone.

\section{Parity Twine on NAQCs}
\label{sec:PTN_NAQC}

So far we have discussed PTNs in a generic setting, considering the DCNOT as a building block. 
However, the DCNOT gate is not a native entangling operation on most available quantum computing platforms.
In particular, on neutral-atom devices the native entangling gate is typically the CZ gate, therefore the DCNOT needs to be transpiled. 
Moreover, many current neutral-atom platforms do not support local addressing for single-qubit rotations in the $XY$ plane, but can only apply them globally.
Therefore, to fully leverage Parity-Twine-based circuit synthesis on NAQCs, it is instructive to follow a co-design approach to optimize PTNs for the underlying hardware capabilities.
To this end, we first briefly review different NAQC architectures, with particular emphasis on how qubit connectivity is established and, in particular, whether atom shuttling is allowed during computation. We then propose different ways to implement PTNs on NAQCs, focusing first on the role of entangling gates before introducing the global rotations.
By construction, all these implementations are unitarily equivalent. 
However, circuit gate counts and depth, as well as the number and distance of potentially required atom shuttling operations, can vary substantially among different physical implementations. To identify the optimal compilation strategy, it is therefore necessary to evaluate them under experimentally realistic parameters with a comprehensive noise model.

\subsection{Neutral-atom architectures}
\label{sec:NAQC_architectures}

Neutral-atom quantum computing architectures can broadly be distinguished by whether the positions of the atoms remain fixed during computation or can be dynamically reconfigured. 
In a \emph{static} architecture atoms are initially arranged in (almost) arbitrary patterns to realize the desired qubit layout, but remain at their fixed positions throughout the computation.
Entangling gates are consequently performed between atoms at fixed positions, and the compilation problem is primarily determined by the available static connectivity and native gate set. In a \emph{mobile} architecture, by contrast, atoms can be transported during computation. Atom transport can be used to bring atoms into or out of interaction regions, implement native physical SWAP operations, dynamically modify the effective qubit connectivity, or to suppress unwanted interactions and crosstalk between non-addressed qubits. Since atom transport and native gate operations operate on substantially different timescales and offer different degrees of parallelization, the tradeoff between connectivity and movement constitutes a nontrivial architectural consideration in NAQCs~\cite{Schmid2023, Saffman2025}.

In this work, we consider neutral-atom platforms that use spatial light modulators (SLMs) to define static trap positions and crossed two-dimensional acousto-optic deflectors (AODs) to provide dynamically movable traps. Atoms can be coherently transferred between AOD- and SLM-generated optical tweezers by adiabatically exchanging their trapping potentials through coordinated intensity ramps~\cite{Barredo2016}, while preserving the qubit state. This capability can be used to prepare the desired geometry before computation in both static and mobile architectures. In mobile architectures, it can additionally be used to dynamically rearrange atoms during computation. 

The distinction between static and mobile architectures is directly relevant to the implementation of PTNs. In this work, we focus primarily on mobile, monolithic architectures, in which atom transport dynamically modifies the relative positions of qubits within a common array. For such architectures, a central compilation objective is to minimize both the number and depth of atom movement operations, since transport contributes substantially to the overall circuit runtime and, consequently, to idling errors. Minimizing movement depth naturally favors parallel transport of multiple atoms whenever possible. Such collective atom transport has been experimentally demonstrated in neutral-atom architectures, including, for example, the implementation of transversal logical operations in quantum error correction experiments~\cite{Bluvstein2024, Rines2026}.

The PTN framework can, in principle, also be extended to zone-based architectures, in which atoms are distributed among distinct computational, storage, and readout zones. In such architectures, the connectivity encoded by a PTN could inform and optimize atom-shuttling strategies by specifying which pairs or groups of atoms should interact, with the corresponding interactions realized through transport between zones. Developing a compilation scheme tailored to such zone-based architectures, however, is beyond the scope of this work.
In particular, the CZSWAP-based implementation discussed below could be a strong candidate for a shuttling-efficient realization on zone-based platforms: since the qubits must be shuttled back to the storage zone after the CZ gate regardless, returning them to the swapped positions instead of their original positions incurs only minimal shuttling overhead.

\begin{figure*}[ht]
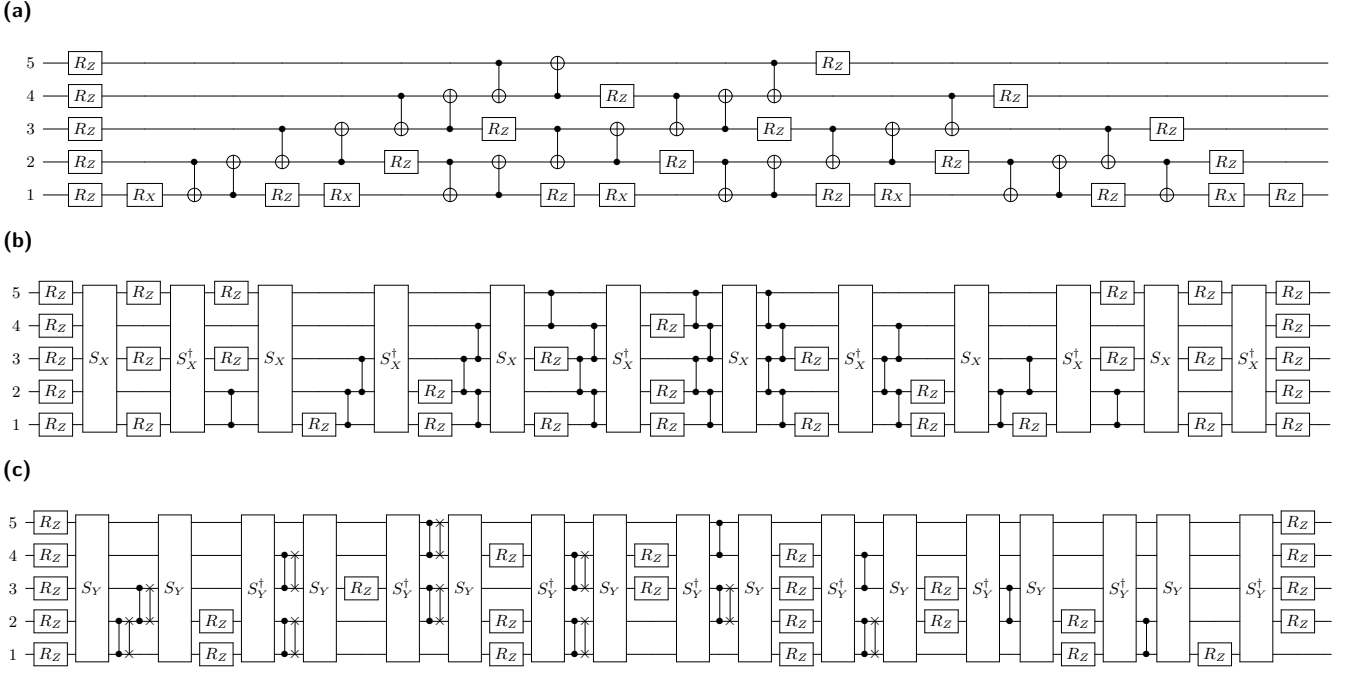

\raggedright
\subfig{(a)}{
\scalebox{0.68}{
    \input{circuits/pt-qft-linear-dcnot}
    }
}\\
\subfig{(b)}{
\scalebox{0.68}{
    \input{circuits/pt-qft-linear-swapless}
    }
}\\
\subfig{(c)}{
\scalebox{0.68}{
    \input{circuits/pt-qft-linear-swapfull}
    }
}
\caption{The quantum Fourier transform on $n=5$ qubits ($\mathrm{QFT}_{5}$) implemented using a Parity Twine Network on a linear nearest-neighbor array, and transpiled into (a) a non-native CNOT gate set, (b) a CZ gate set with global $R_X$ rotations, and (c) a CZSWAP gate set with global $R_Y$ rotations. The CZSWAP operation consists of a CZ gate followed by a physical SWAP of the underlying atoms, enabled by atom shuttling in a shuttling-capable architecture. For visual clarity, the rotation angles of the $R_Z$ gates, which differ across qubits, are omitted from the circuit diagrams.}
\label{fig:qft_circuits}
\end{figure*}

\subsection{Entangling gates}
\label{sec:entangling gates}

Within NAQCs, three primary decomposition strategies for the DCNOT building block can be considered, as shown in Fig.~\ref{fig:ptc}(b):
\begin{enumerate}[label=(\roman*)]
\item Decomposition into two CZ gates
\begin{equation*}
\mathrm{DCNOT} = \left(I \otimes H \right) \mathrm{CZ} \left(H \otimes H \right) \mathrm{CZ} \left(H \otimes I \right);
\end{equation*}

\item Mapping to an iSWAP gate
\begin{equation*}
\mathrm{DCNOT} = \left(H \otimes I \right) \mathrm{iSWAP} \left(S^{\dagger} \otimes S^{\dagger} H\right);
\end{equation*}

\item Realization as a combined CZ--SWAP (CZSWAP) operation, i.e., a CZ gate followed by a SWAP operation on the same two qubits
\begin{equation*}
\mathrm{DCNOT} = \left(H \otimes I \right) \mathrm{CZSWAP} \left(I \otimes H \right).
\end{equation*}

\end{enumerate}

Decomposition (i) is the standard approach and preserves the two-qubit gate count of the original PTN construction. 
Decomposition (ii)---which reduces two CNOTs to a single iSWAP---is, in principle, the most efficient in terms of two-qubit gate count. 
However, the iSWAP gate is not readily available on most current neutral-atom platforms, but schemes to realize high-fidelity iSWAP gates have recently been proposed~\cite{Bergonzoni2026, Ildefonso2026}.
Decomposition (iii), involving a CZSWAP operation, may initially appear suboptimal, as it introduces an additional, usually costly SWAP operation alongside the CZ gate.
Nevertheless, as detailed below, in mobile, monolithic architectures the effective cost of a displacement-based SWAP can be negligible when it is performed after a CZ gate, rendering this approach highly competitive.

To show this, it is useful to briefly review how a Rydberg-mediated CZ gate between two atoms, A and B, is typically implemented in such architectures:
\begin{enumerate}[label=(\arabic*)]   
    \item Atom A is transferred from a static optical tweezer array (SLM) into a mobile tweezer (AOD) and transported into proximity with atom B, such that both reside within the Rydberg blockade radius;
    \item Atom A is then released into a static trap, and the entangling operation is performed;
    \item Atom A is recaptured by the mobile tweezer, transported back to its original position, and reloaded into the static array.
\end{enumerate}
In total, such a CZ gate requires two \emph{big move} transport operations and four transfer operations between static and mobile tweezers. We define a \emph{big move} as the transport from the initial distance of the atoms to the shorter interaction distance for performing the CZ gate operation.
While Rydberg-mediated CZ gates are typically performed with the atoms held in static traps, we assume throughout this work that inter-tweezer CZ gates are also allowed. In this case, step (2) can be performed without transferring atom A into the SLM array, such that the tweezer transfer count required for a single CZ gate is reduced from four to two.
Consequently, a DCNOT implemented via two independently performed CZ gates requires four transfer operations and four big moves in total. 
Also note that if local gates can be applied while the atoms remain in proximity, the two CZ gates in decomposition~(i) can be applied consecutively without moving the atoms apart in between; this case is not considered in the mobile architecture assumed in this work.

The CZSWAP gate can be performed with only a minimal modification of step (3): Instead of returning atom A to its original position, atom B is transported back to atom A's initial position, while atom A remains at its new position.
This does not change the big move count compared to the CZ gate, and the only required additional transport is a short \emph{offset move} needed to exchange the atomic positions, which is negligible compared with the big move~\cite{Gao2025}.
However, since AOD arrays cannot cross during the physical SWAP operation, for the CZSWAP it is always necessary to drop atom A into the static SLM array and to pick up atom B from this array. Therefore, it requires four transfer operations.
Consequently, in such architectures, the CZSWAP construction can achieve higher overall fidelity than a decomposition into two CZ gates by reducing both the number of CZ gates and the total shuttling time.

\subsection{Global rotations in the $XY$ plane}
\label{sec:global xy rotation}
In many state-of-the-art NAQC architectures, rotations around an axis in the $XY$ plane are restricted to global rotations $\mathrm{GR}_{\alpha}(\theta)$ acting on the entire atom array (or a defined subset);
here $\alpha$ denotes the rotation axis in the $XY$ plane and $\theta$ is the rotation angle.
Under these constraints, arbitrary single-qubit rotations can be synthesized by interleaving three layers of local $R_Z$ rotations with two global $XY$ rotations~\cite{Nottingham2023}. 
%While this decomposition allows arbitrary rotations on all qubits to be executed simultaneously by tailoring the local $R_Z$phases, it also means that global $XY$ pulses are unavoidable even when targeting only a small subset of qubits. Consequently, 
Therefore, we additionally optimize our PTNs around global $XY$ rotations for both the CZ-only and CZSWAP decompositions.

To illustrate this, consider the QFT implemented with PTNs [see Fig.~\ref{fig:qft_circuits}].
In the CZ-only decomposition, Hadamard gates must be inserted before, between, and after the two CZ gates [cf.~Fig.~\ref{fig:ptc}(b)].
While the structure of the Twine chains allows some of these to cancel, for example the Hadamards situated between pairs of CZ gates remain and must be decomposed into global $XY$ rotations.
Naively, this would require two global rotation blocks per Hadamard layer.
However, the structured circuit of the PTNs on the LNN chain enables us to reduce this overhead by nearly half, requiring only one global $XY$ rotation per Hadamard layer, plus two additional layers at the start and end of the circuit [see Fig.~\ref{fig:qft_circuits}(b)].
For these operations, we specifically choose alternating global $\pm \pi/2$ $X$ rotations, $S_X \equiv \mathrm{GR}_X(\pi/2)$ and $S_X^\dagger \equiv \mathrm{GR}_X(-\pi/2)$.
When the timing of these global rotations is synchronized, this alternating structure can act like a dynamical decoupling sequence, effectively suppressing hardware errors.
Yet, this choice remains flexible; by adjusting the initial and final rotation layers alongside specific internal $R_Z$ angles, the circuit can also be compiled using uniform $S_X$ blocks.

On the other hand, the CZSWAP-based implementation of the QFT consists of layers of entangling CZSWAP and CZ gates interleaved with single-qubit $X$ rotations. These rotations can be globalized using the simplified decomposition
\begin{equation}
R_X^{(i)}(\theta) = S_Y^\dagger R_Z^{(i)}(\theta) S_Y,
\end{equation}
where ${S_Y \equiv \mathrm{GR}_Y(\pi/2)}$ and ${S_Y^\dagger \equiv \mathrm{GR}_Y(-\pi/2)}$ [see Fig.~\ref{fig:qft_circuits}(c)]. Note that QFT circuits based on the iSWAP decomposition are equivalent to those based on the CZSWAP decomposition, up to additional $S^\dagger$ gates, since
\begin{equation}
\mathrm{CZSWAP} = \left(S^\dagger \otimes S^\dagger\right)\mathrm{iSWAP}.
\end{equation}
Nevertheless, the compiled circuit structure and the $S_Y$ gate count remain unchanged. We therefore focus on decompositions (i) and (iii) throughout this paper.

\begin{table}[!b]
\centering
\caption{Parameters used for the fidelity estimates throughout this work unless specified otherwise.}
\label{tab:params}
\setlength{\tabcolsep}{6pt}
\footnotesize
\begin{tabular}{llr}
\toprule
\textbf{Parameter} & \textbf{Description} & \textbf{Value} \\
\midrule
$f_{\mathrm{CZ}}$      & CZ-gate fidelity             & $99.5\,\%$ \\
$f_{\mathrm{GR}}$      & GR-gate fidelity             & $100.0\,\%$ \\
$f_{\mathrm{exc}}$     & Rydberg excitation fidelity  & $99.75\,\%$ \\
$f_{\mathrm{trans}}$   & Trap transfer fidelity       & $99.9\,\%$ \\
\midrule
$T_2$                  & Atom coherence time               & $1.5\,\mathrm{s}$ \\
$T_{\mathrm{trans}}$   & Trap transfer time           & $1.5\,\mathrm{\mu s}$ \\
$T_{\mathrm{CZ}}$      & Rydberg pulse time           & $0.36\,\mathrm{\mu s}$ \\
$a$                    & effective atom acceleration         & $2750\,\mathrm{m/s^2}$ \\
\midrule
$d_{\mathrm{tw}}$      & Interatomic spacing        & $15\,\mathrm{\mu m}$ \\
$d_{\mathrm{int}}$      & CZ interaction distance      & $3\,\mathrm{\mu m}$ \\
$d_{\mathrm{big}}$     & Big-move distance           & $12\,\mathrm{\mu m}$ \\
$d_{\mathrm{offset}}$  & Offset-move distance         & $2\,\mathrm{\mu m}$ \\
\bottomrule
\end{tabular}
\end{table}

\subsection{Fidelity estimation and experimental parameters}

To evaluate the performance of the different PTN-based implementations on NAQCs and to benchmark them against alternative circuit proposals~\cite{Gao2025} and NA-specific compilers~\cite{Tan2025, Wang2024} we estimate the total fidelity $f$ of a given quantum circuit (transpiled to the native gate set) by~\cite{Schmid2024, Tan2025, Gao2025}

\begin{equation}
\label{eq:circuit_fidelity}
\begin{split}
f =& (f_{\mathrm{GR}})^{N_{\mathrm{GR}}} \cdot (f_{\mathrm{CZ}})^{N_{\mathrm{CZ}}} \cdot
(f_{\mathrm{exc}})^{N_q S - 2N_{\mathrm{CZ}}}
\\
& \cdot (f_{\mathrm{trans}})^{N_{\mathrm{trans}}} \cdot \prod_{j=1}^{N_q} \left(1 - \frac{T_{\mathrm{idle}}^{j}}{T_2}\right).
\end{split}
\end{equation}
Table~\ref{tab:params} details all relevant experimental parameters used in Eq.~\eqref{eq:circuit_fidelity} and throughout this paper, unless specified otherwise.
The first two terms describe the fidelity from performing the global rotation gates and the entangling CZ gates, respectively. Here, $N_{\mathrm{GR}}$ and $N_{\mathrm{CZ}}$ denote the numbers of global rotations and CZ gates.
It is worth mentioning that we assume unit fidelity for global rotations, ${f_{\mathrm{GR}}=1.0}$. 
This assumption should be understood as a normalization rather than an assumption of error-free single-qubit gates. 
Since reliable fidelity data for global single-qubit rotations are not available for the non-PTN implementations considered here as comparisons, which rely on local single-qubit operations, we factor out single-qubit gate errors and focus the fidelity analysis on the entangling operations and atom transport. 
In this sense, the reported fidelities can be viewed as normalized with respect to the total fidelity contribution of single-qubit gates.

Crosstalk arising from unintended Rydberg excitations is modeled by the third term in Eq.~\eqref{eq:circuit_fidelity}.
Here, $f_{\rm exc}$ denotes the idling fidelity associated with the unintentional excitation of idling atoms to the Rydberg state, while ${N_q S-2N_{\mathrm{CZ}}}$ quantifies the number of idling qubits during the execution of CZ gates across all Rydberg stages, with each stage affecting all $N_q$ qubits.
$S$ denotes the number of Rydberg stages, with each stage defined as a time interval during which the Rydberg laser is applied to induce one or more parallel Rydberg-mediated CZ gates.

The fourth term captures the fidelity associated with atom transfers between dynamic AOD and static SLM traps. The total number of transfer operations is given by
${N_{\mathrm{trans}} = 2N_{\mathrm{CZ}} + 4N_{\mathrm{CZSWAP}}}$, as discussed in Sec.~\ref{sec:entangling gates}.

Finally, idling errors are described by the last term in Eq.~\eqref{eq:circuit_fidelity}, where $T_{\mathrm{idle}}^{q}$ denotes the total idle time accumulated by qubit $q$.
Throughout this work, we assume that dynamical decoupling sequences are applied during all idle periods and atom transports to suppress dephasing~\cite{Manetsch2025}.
Consequently, the $T_2$ value reported in Table~\ref{tab:params} represents the effective coherence time achieved under these sequences~\cite{Evered2023}.
Because the total shuttling and idle times remain well below $T_2$, the residual decoherence from idling is approximated to first order as a linear decay.

We assume that errors arising during atom shuttling are fully captured by the idling errors. To estimate the shuttling durations, we use the following geometric parameters. We assume a Rydberg blockade radius of ${6\,\mu\mathrm{m}}$ and an interatomic spacing of ${d_{\rm tw}=15\,\mu\mathrm{m}}$ to suppress crosstalk. 
For high-fidelity CZ gates, atoms are brought to a separation of ${d_{\rm int}=3\,\mu\mathrm{m}}$, well within the blockade radius. This corresponds to a displacement of ${d_{\rm big}=12\,\mu\mathrm{m}}$ for each of the two big-move shuttling operations. 
For the implementation of a CZSWAP operation, we require two big-move operations of the same distance, together with four additional offset moves of ${d_{\rm offset}=2\,\mu\mathrm{m}}$ each to prevent the AOD arrays from crossing [see App.~\ref{app:czswap} for further details on the CZSWAP implementation].
The transport time is estimated using the scaling ${t=\sqrt{d/a}}$, where $d$ is the displacement associated with the corresponding CZ or CZSWAP operation and $a$ is the effective atom acceleration calibrated to the experimentally demonstrated transport time of $200\,\mu\mathrm{s}$ over $110\,\mu\mathrm{m}$ reported in Ref.~\cite{Bluvstein2022}.

Although Eq.~\eqref{eq:circuit_fidelity} provides only a first-order fidelity approximation, it facilitates direct benchmarking against alternative proposals and NA-specific compilers using a concise set of experimental parameters.
Crucially, because it relies exclusively on circuit-level metrics, this model scales to significantly larger qubit numbers than would be feasible with full circuit simulations.

\section{Algorithmic demonstrations}\label{sec:Applications}

Having established realizations of the Parity Twine framework for neutral-atom platforms and discussed our performance metric, we now discuss several algorithmic use-cases in detail and benchmark different realizations against previously suggested implementations and publicly available neutral-atom compilers.

\subsection{Exact QFT}
\label{sec:qft}

The first use-case we consider is the exact QFT algorithm implemented with the PTN-based approaches described above.
Given a PTN, the implementation of the QFT involves the addition of local rotations around X and Z throughout the entire circuit [see Fig.~\ref{fig:qft_circuits}(a)].
While the single-qubit $R_Z$ rotations can be readily included, the $R_X$ rotations must be performed through global operations, as discussed in Sec.~\ref{sec:global xy rotation}.
However, no additional global gates are required, since these $R_X$ operations can be absorbed directly into the global pulses arising from the transpilation of the Twine chains, maintaining the optimality of the circuit.

Figure~\ref{fig:qft_circuits} illustrates PTN-based QFT circuits ($\mathrm{QFT}_n$) for a representative small system of ${n=5}$ qubits in a linear layout comparing the basic implementation based on DCNOT with local rotations [Fig.~\ref{fig:qft_circuits}(a)], the native CZ-only decomposition with global $XY$ rotations [Fig.~\ref{fig:qft_circuits}(b)], and the native CZSWAP decomposition with global $XY$ rotations [Fig.~\ref{fig:qft_circuits}(c)]. 

%Since PTNs can be embedded into hardware layouts with different connectivities and implemented using different native gate decompositions, an important question is how these choices influence the trade-off between gate count, circuit depth, and atom movement. We first compare the resource requirements of the CZSWAP-based QFT across different qubit layouts and then estimate the resulting circuit fidelities, contrasting the CZSWAP implementation with the CZ-only realization.

\subsubsection*{PTN benchmarking}

\begin{figure}[!t]
\centering
\begin{tabular}{c}
  \subfig{(a)}{
\includegraphics[width=\columnwidth]{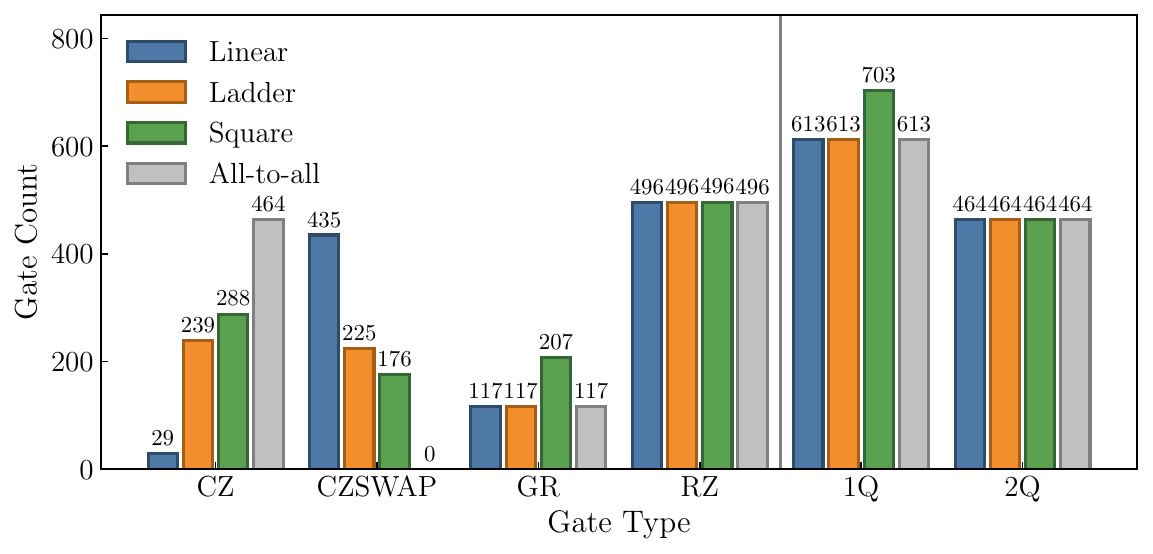}
  } \\
  \subfig{(b)}{
\includegraphics[width=\columnwidth]{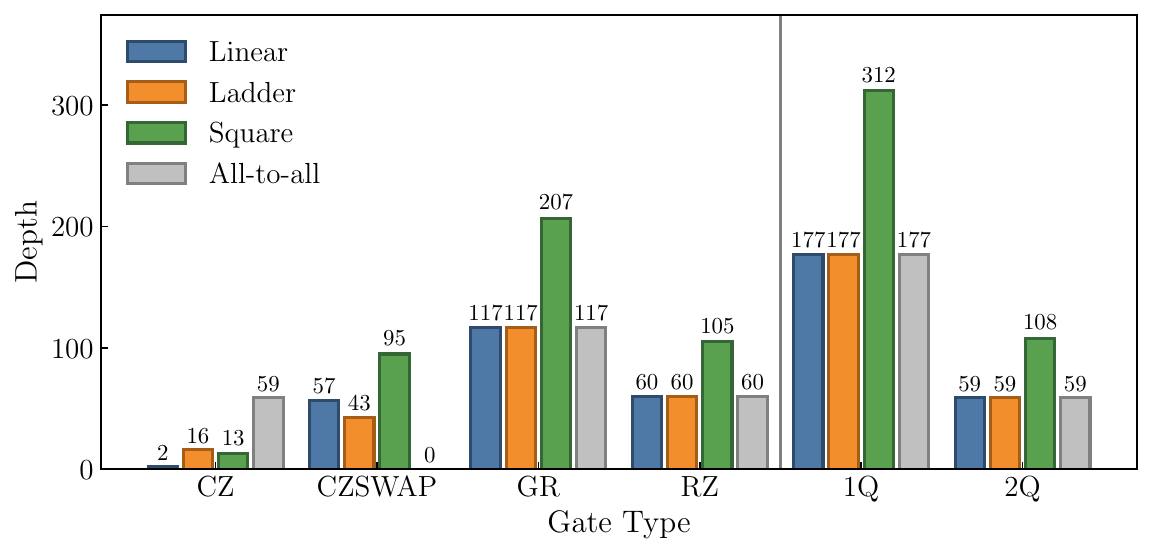}
  }
\end{tabular}
\caption{Per-gate-type and total single- and two-qubit (a) gate counts and (b) gate depths for the $\mathrm{QFT}_{30}$ implemented using a CZSWAP-based Parity Twine network across different hardware layouts. Here, the 1Q category includes global rotations (GR) and local RZ rotations, while the 2Q category includes CZ and CZSWAP gates.
}
\label{fig:gate_count_depth}
\end{figure}

Figure~\ref{fig:qft_circuits}(c) illustrates the CZSWAP implementation on a linear layout, which utilizes CZSWAP gates throughout the Twine chains and employs CZ gates exclusively for the final decoding stage.
Conversely, the enhanced connectivity of the ladder layout---as an example---minimizes the number of CZSWAP operations generated during DCNOT conversion.
Despite this reduction in CZSWAP gates, the combined two-qubit gate count, treating both CZ and CZSWAP equally as single two-qubit operations, remains identical across both layouts\footnote{Note that this definition differs from the original PTN formulation in Ref.~\cite{Dreier2025} where each CNOT is counted individually.}.

In Fig.~\ref{fig:gate_count_depth}, we present a summary of the gate counts and circuit depths of transpiled $\mathrm{QFT}_{30}$ circuits implemented via PTNs using the CZSWAP decomposition for four layouts: linear, ladder, square, and all-to-all.
Figure~\ref{fig:gate_count_depth}(a) confirms that the total number of two-qubit gates remains invariant across the different layouts, while the number of CZSWAP operations decreases with increasing connectivity, as discussed. The all-to-all connectivity limit provides a useful reference point beyond the architectures considered here: in this limit, no backbone of DCNOT gates is required to transport parity labels, and hence no CZSWAP operations are needed.
Instead, long-range CZ gates alone can propagate parity label information throughout the entire system~\cite{Dreier2025}. 
Although the total two-qubit gate count remains identical across all layouts, the square layout requires more single-qubit gates than the linear and ladder configurations.
This overhead arises from an increased number of global rotations necessitated by a larger two-qubit gate depth, as detailed below.

Figure~\ref{fig:gate_count_depth}(b) shows the corresponding circuit depth, resolved by gate type. 
In particular, entangling layers are separated into those consisting exclusively of CZ gates and those containing at least one CZSWAP operation. 
The linear and ladder layouts show identical total two-qubit gate depth, although the ladder layout reduces the CZSWAP depth at the expense of CZ depth compared to the linear layout.
In contrast, while the increased connectivity of the square layout reduces the CZSWAP gate count [cf.~Fig.~\ref{fig:gate_count_depth}(a)], it exhibits an increased CZSWAP and overall two-qubit gate depth compared to the linear and ladder layouts.
This increased two-qubit gate depth also drives up the number and depth of global rotation gates, because fewer single-qubit rotations are applied per global rotation gate.
These results indicate that parallelization is generally more constrained in layouts with higher connectivity~\cite{Dreier2025}, as reflected by the increased depth contributions from single-qubit gates and CZSWAP operations. 
Notably, the ladder architecture serves as a counterexample, maintaining the same depth for single- and two-qubit gates as the linear layout.
This is possible because it utilizes two parallel backbones, instead of the single backbone used in the linear layout, as discussed in Sec.~\ref{sec:PTN}.

These circuit metrics directly dictate physical performance by determining the required gate counts, shuttling operations, and overall execution time.
To quantify this impact, we now estimate the fidelity of PTN-based $\mathrm{QFT}_{30}$ circuits across different layouts using Eq.~\eqref{eq:circuit_fidelity}, benchmarking our results against state-of-the-art compilers designed for neutral-atom platforms with mid-circuit rearrangement and previously proposed \textit{optimal} implementations of the QFT for neutral-atom devices from Ref.~\cite{Gao2025}.
More specifically, we consider the Enola~\cite{Tan2025} and Atomique~\cite{Wang2024} compilers: Enola minimizes two-qubit gate depth while enabling dynamic reconfiguration after each Rydberg stage, whereas Atomique exploits parallel atom transport via multiple AOD arrays to reduce transfers between AOD and SLM traps, potentially at the cost of adding ancilla atoms to the circuit.

\begin{table}[!b]
\centering
\setlength{\tabcolsep}{3pt}
\footnotesize
\caption{Benchmarking of compilation strategies for the $\mathrm{QFT}_{30}$ circuit across different hardware layouts.}
\begin{tabular}{llcccccc}
\toprule
\textbf{Method} & \textbf{Layout} & $\bm{N}_{\mathbf{CZ}}$ & $\bm{N_q}$ & $\bm{S}$ & $\bm{N}_{\mathbf{trans}}$ & \makecell{\textbf{Big}\\\textbf{move}} & \makecell{\textbf{Offset}\\\textbf{move}} \\
\midrule
PTN & linear   & \textbf{464} & 30 & \textbf{59}  & 1798 & \textbf{117} & 228 \\
PTN & ladder   & \textbf{464} & 30 & \textbf{59}  & 1378 & \textbf{117} & \textbf{172} \\
PTN & square & \textbf{464} & 30 & 108 & 1236 & 215 & 380 \\
Gao \cite{Gao2025} & linear   & 870 & 30 & 114 & 3478 & 227 & 681 \\
Gao \cite{Gao2025} & square   & 870 & 30 & 114 & 3478 & 227 & 1413 \\
Enola \cite{Tan2025} & arbitrary     & 870 & 30 & 114 & 3478 & 813 & 4681 \\
Atomique \cite{Wang2024} & arbitrary  & 954 & 43 & 466 & \textbf{0}     & 237 & 229 \\
\bottomrule
\end{tabular}
\label{tab:comparison}
\end{table}

\begin{figure}[!t]
\centering
\includegraphics[width=\columnwidth]{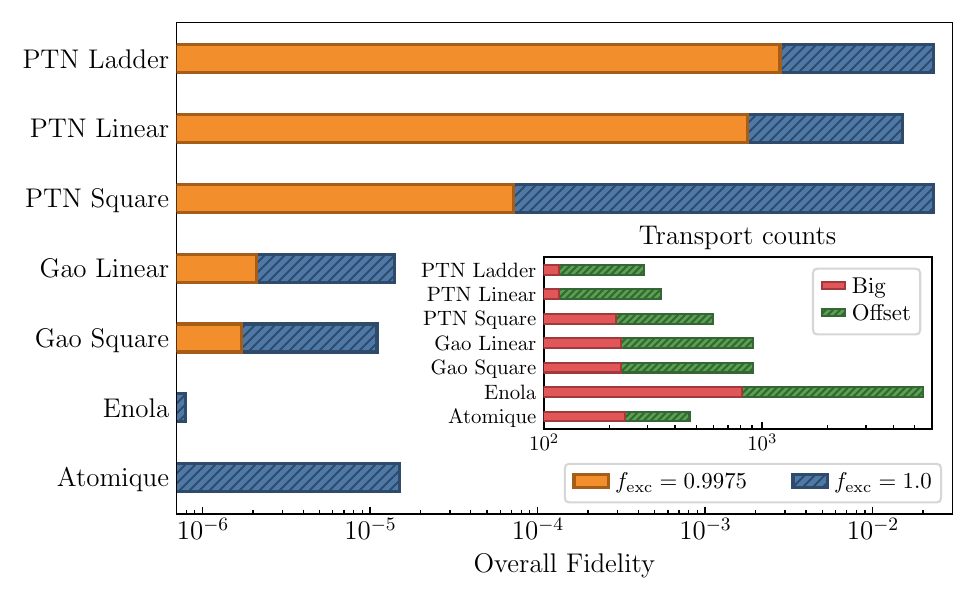}
\caption{Estimated fidelity of the $\mathrm{QFT}_{30}$ circuit for different compilation methods. 
Blue bars show results for vanishing Rydberg crosstalk ($f_{\rm exc}=1$), while orange bars consider a $0.25\%$ crosstalk error ($f_{\rm exc}=99.75\%$).
The inset shows the numbers of big moves (red bars) and offset moves (green bars) required for both CZ and CZSWAP gate implementations.
Data for all non-PTN methods are taken from Ref.~\cite{Gao2025}.}
\label{fig:fidelity}
\end{figure}

Table~\ref{tab:comparison} lists the relevant operational parameters required for the fidelity estimation for the different considered approaches. 
An uncompiled $\mathrm{QFT}_{n}$ circuit requires ${n(n-1)/2}$ controlled-phase gates, which can be decomposed into ${N_{\rm CZ} = {n(n-1)}}$ CZ gates, as in Gao’s approach~\cite{Gao2025}.
Enola requires the same number of CZ gates, but requires substantially more transport (\textit{big move} and \textit{offset move}) operations than Gao [see also inset in Fig.~\ref{fig:fidelity}].
Atomique reduces both large and offset movements by eliminating transfer operations through the use of multiple AOD arrays\footnote{For the Atomique results reported in Ref.~\cite{Gao2025}, two AOD arrays are used throughout. Thus, the reported performance corresponds to a fixed AOD-array count of two.}. However, this comes at the cost of additional ancilla qubits and gate-based SWAP operations, resulting in the largest CZ count among all methods. 
In contrast, the Parity Twine framework reduces the number of CZ operations (CZ and CZSWAP gates) to ${N_{\rm CZ}={n(n-1)/2}}$ for the Twine chains, supplemented by a decoding sequence of ${n-1}$ CZ gates, yielding a total of ${n(n+1)/2 -1}$ [see more details in App.~\ref{app:resource_scaling}]. 
This constitutes a significant reduction in the two-qubit gate count compared to the other methods, approaching a factor of 2 for large $n$. 
Furthermore, this decrease translates into substantially fewer atom transport and trap transfer operations compared to the competing approaches.

\begin{figure*}[!ht]
\centering
\begin{tabular}{c}
  \subfig{(a)}{
\includegraphics[height=0.24\textheight]{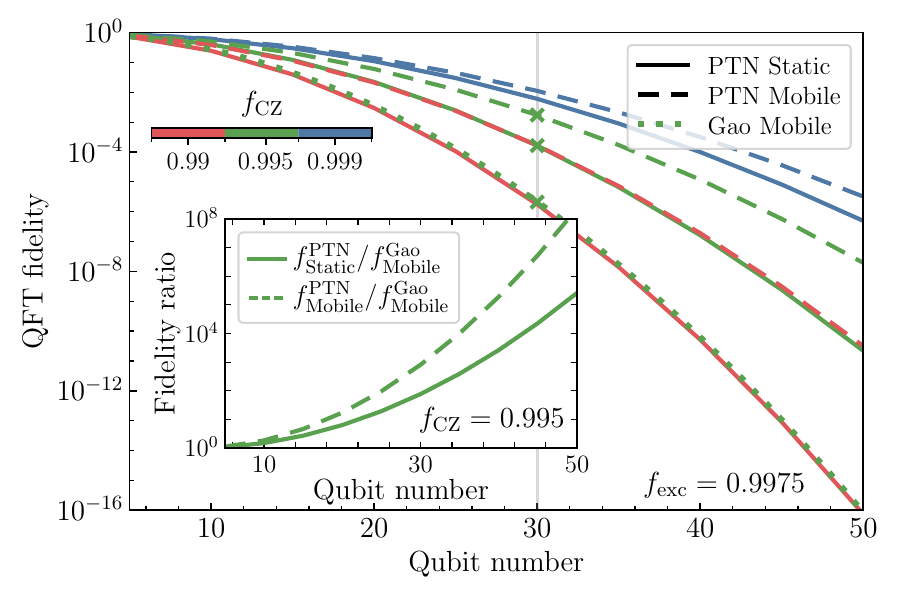}
  } 
  \subfig{(b)}{
\includegraphics[height=0.24\textheight]{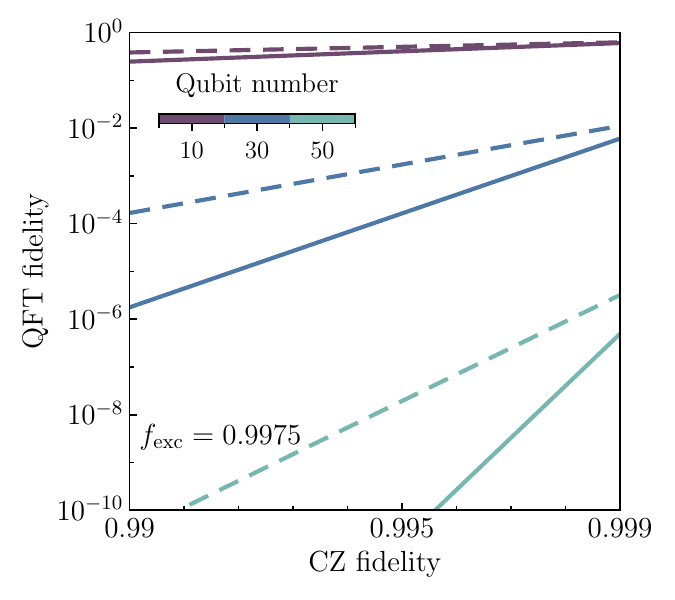}
  }
  \subfig{(c)}{
\includegraphics[height=0.24\textheight]{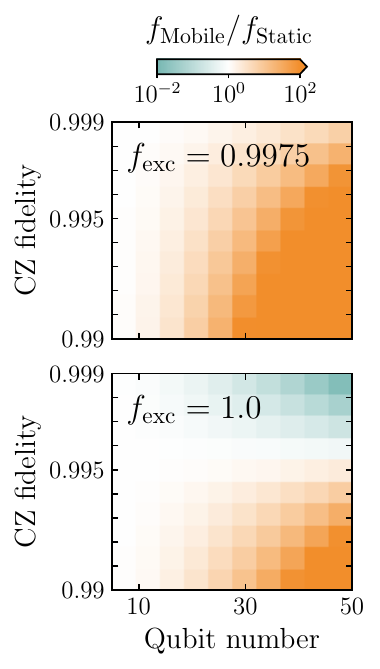}
  }
\end{tabular}
\caption{Estimated QFT fidelity on a linear neutral-atom array using static (solid lines) and mobile (dashed lines) architectures as a function of (a) the number of qubits and (b) the CZ-gate fidelity. The dotted lines in panel (a) show the best previously reported QFT fidelities at ${f_\mathrm{CZ}=0.995}$ obtained using Gao's linear compilation~\cite{Gao2025}. Crosses mark the corresponding QFT fidelities at 30 qubits for the three compilation approaches for a CZ gate fidelity of $f_{\rm CZ}=99.5\%$. The inset shows the fidelity improvements of PTN-based implementations relative to Gao's approach. (c) Ratio of the mobile to static PTN-based QFT fidelity in the presence (upper panel) and absence (lower panel) of Rydberg crosstalk.
}
\label{fig:static_vs_mobile}
\end{figure*}

Driven by this combined reduction in two-qubit gate count, shuttling overhead, and the number of Rydberg stages, our Twine-based QFT achieves an estimated overall circuit fidelity up to three orders of magnitude higher than the best competing implementation, as shown in Fig.~\ref{fig:fidelity}.
In the absence of Rydberg crosstalk $f_{\rm exc}=1$ (blue bars), it shows similar fidelities across different qubit layouts. 
When crosstalk is included (orange bars), the ladder layout achieves the highest fidelity among the three PTN layouts. 
In contrast, the square layout is significantly more affected by crosstalk due to its larger two-qubit gate depth and, consequently, the higher number of Rydberg stages. 
Both the linear and ladder PTN layouts demonstrate exceptional performance regardless of crosstalk.
Still, the ladder layout achieves a higher fidelity than the linear configuration by replacing a portion of the CZSWAP gates with standard CZ gates.
Although this substitution maintains the same total number of entangling operations, it significantly reduces the required SLM-AOD transfers $N_{\rm trans}$ and offset moves, thereby yielding the fidelity advantage.

Ultimately, these results demonstrate the high efficiency of the Parity Twine formalism also for NAQCs.
Moreover, they highlight its adaptability across various qubit layouts, enabling the co-design of fidelity-optimal configurations tailored to specific hardware parameters and competing operational constraints. 

\subsubsection*{Mobile vs static architectures}

Having established the high efficiency of PTN-based implementations of the QFT using the CZSWAP decomposition for mobile NAQC architectures, we now compare this implementation with the CZ-only PTN implementation on a static NAQC architecture without mid-circuit atom transport, see Sec.~\ref{sec:NAQC_architectures}.
Figure~\ref{fig:static_vs_mobile} shows this comparison for the QFT on a linear qubit layout.
In the static architecture, the SWAP-free compilation requires only nearest-neighbor CZ gates and therefore no atom shuttling. 
Our fidelity estimates nevertheless include a residual crosstalk infidelity of $0.25\%$ per Rydberg excitation ($f_{\rm exc} = 0.9975$) while again assuming noiseless single-qubit operations.

The fidelity estimates are plotted as a function of qubit number $n$ in Fig.~\ref{fig:static_vs_mobile}(a) for the static (full lines) and mobile (dashed lines) architectures, where line color encodes different CZ gate fidelities.
For comparison, we also include the previous best reported QFT fidelities based on Gao's compilation strategy~\cite{Gao2025} for linear arrays (dotted lines). 
Both PTN approaches exhibit an increasingly pronounced fidelity advantage over Gao's approach as the system size grows [see also inset in Fig.~\ref{fig:static_vs_mobile}(a)]. 
For example, at ${n=30}$ and a CZ gate fidelity of 99.5\%, the estimated circuit fidelity improves by approximately two orders of magnitude for the static implementation and by three orders of magnitude for the mobile implementation. Remarkably, the PTN-based implementations achieve substantially higher fidelities than Gao's compilation even at a CZ-gate error rate twice as large, highlighting how the reduced entangling-gate count and circuit depth of the PTN approach mitigate the impact of imperfect CZ gates.

Interestingly, the CZSWAP-based (mobile) implementation consistently outperforms the SWAP-free (static) realization in the presence of Rydberg crosstalk, even for CZ-gate fidelities approaching $0.999$ [see also Fig.~\ref{fig:static_vs_mobile}(b)]. 
This occurs because, in the CZ-only static implementation, each CZSWAP required by the mobile approach is replaced by two CZ gates, resulting in an almost twofold increase in the total number of CZ gates.
As shown in the top panel of Fig.~\ref{fig:static_vs_mobile}(c), this mobile advantage becomes even more pronounced at lower CZ-gate fidelities and larger system sizes.
However, when crosstalk is neglected (${f_{\rm exc} = 1.0}$), the SWAP-free implementation becomes advantageous once the CZ-gate fidelity exceeds ${f_{\mathrm{CZ}} \approx 0.996}$ [see bottom panel in Fig.~\ref{fig:static_vs_mobile}(c)].
At this threshold, the infidelity introduced by atom transfers and time-intensive shuttling in the mobile approach outweighs the penalty of the additional CZ gates required by the static approach.
This crossover clearly illustrates that the optimal compilation strategy is highly sensitive to the underlying hardware characteristics. 
We emphasize, however, that the fidelity estimates presented here serve primarily as performance indicators; exact crossover points will depend on the specific capabilities and limitations of the target neutral-atom platform.

In summary, we have shown that PTNs constitute a highly efficient approach to synthesize circuits with high-interaction density, such as the QFT, for NAQCs.
Furthermore, PTNs span the continuum between fully static, fully mobile, and hybrid neutral-atom architectures. 
Depending on the available movement primitives, the native entangling operations, and the possible qubit layouts, the same Parity Twine construction can automatically trade atom transport for additional entangling gates or vice versa, ensuring high-fidelity circuit execution.

\subsection{Optimistic QFT}
\label{sec:oqft}
The exact QFT requires controlled-phase gates in the textbook circuit representation~\cite{NielsenChuang2010}, or single-qubit phase gates in the PTN representation, whose rotation angles shrink exponentially, ${\theta_k = \pi/2^k}$ for ${k = 2, \ldots, n}$, leading to $\mathcal{O}(n^2)$ gates and linear circuit depth. 
It has long been recognized that the smallest of these rotations barely affect the output state and can be safely dropped~\cite{Coppersmith2002}. Keeping only ${m=\mathcal{O}(\log(n/\epsilon))}$ bits of each qubit's binary phase, corresponding to retaining rotations up to $k\leq m$, yields the approximate QFT (AQFT), which approximates the exact QFT within a fixed error $\epsilon$, uniformly over the Hilbert space, while reducing the gate count to ${\mathcal{O}(nm)\simeq\mathcal{O}(n\log(n/\epsilon))}$~\cite{Coppersmith2002}\footnote{This scaling refers to the standard unitary implementation without ancillary qubits or mid-circuit measurements; allowing additional resources or more general computational models can yield improved asymptotic scalings.}.

Building on this idea, Ref.~\cite{Kahanamokumeyer2025} introduced the optimistic QFT (OQFT), which further exploits the approximate commutativity of AQFT blocks to reduce circuit depth to $\mathcal{O}\left(\log (n/\epsilon)\right)$ without changing the AQFT gate-count scaling. 
The OQFT closely approximates the exact QFT for the vast majority of input basis states, while larger errors are confined to a small subset of states, comprising only an $\mathcal{O}(\epsilon)$ fraction of the Hilbert space~\cite{Kahanamokumeyer2025}. 
Remarkably, uniform accuracy across all computational-basis states is not necessary for the OQFT to remain useful \cite{Kahanamokumeyer2025}: by concentrating its errors on a small subset of exceptional states, it can retain high fidelity on most of the input space, suggesting that its average-case performance may be sufficient for quantum algorithms relying on the QFT, as has been demonstrated for AQFTs in applications such as phase estimation and period finding \cite{Linden2022}.

\begin{figure*}[t!]
\begin{tabular}{c}
  \subfig{(a)}{
\includegraphics[width=0.56\textwidth]{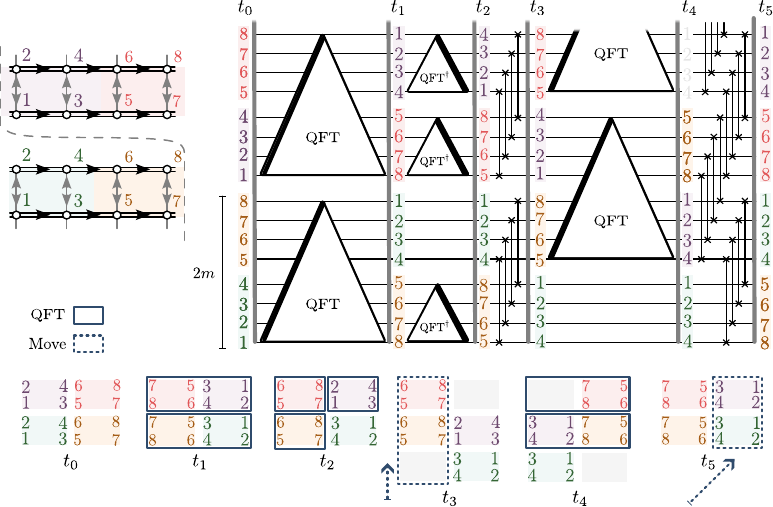}
  }
  \subfig{(b)}{
\includegraphics[width=0.42\textwidth]{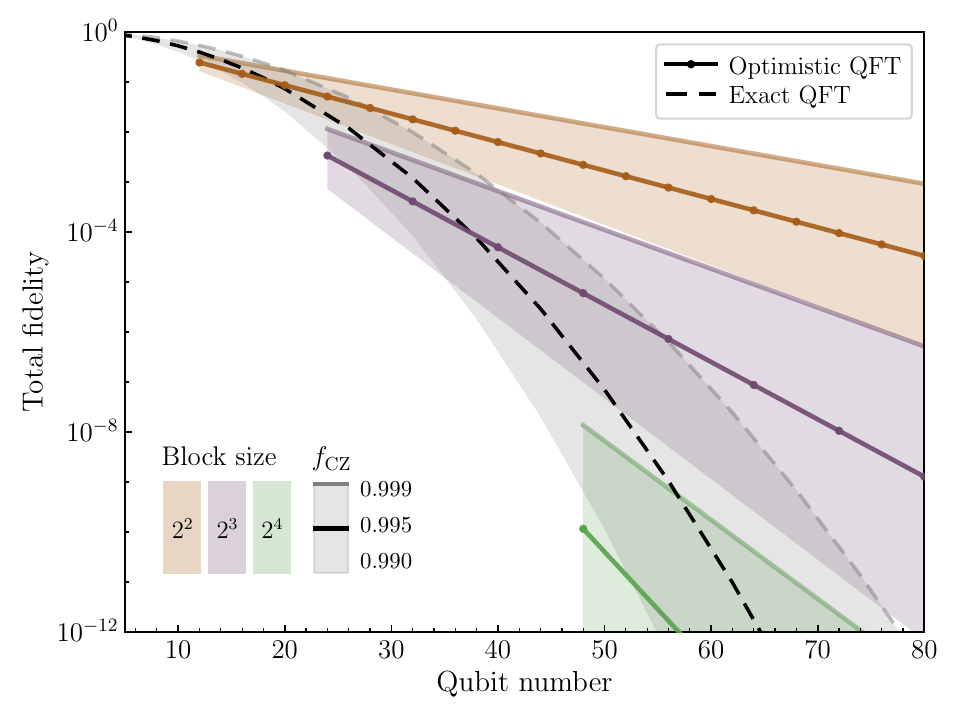}
  }
\end{tabular}
\caption{(a) Optimistic QFT constructed from exact QFT subcircuits of sizes $m$ and $2m$ using the ladder Parity Twine network (triangles), embedded within a larger square grid (Twine path shown). The qubit ordering evolves at each time step $t_i$, necessitating nearest-neighbor block exchanges at $t_2$ and $t_4$ to restore the order at the final time $t_5$, up to a left-right reversal. As illustrated for four blocks, these exchanges naturally map to translations of the first column of blocks in a mobile neutral-atom architecture (bottom plots).
(b) Estimated total fidelity of the PTN-based optimistic QFT as a function of the total number of qubits, $n$, for different QFT block sizes, $m$, assuming a mobile neutral-atom architecture. Solid lines correspond to the optimistic QFT, while the dashed line shows the exact QFT as a comparison. The translucent bands indicate the fidelity range obtained for CZ gate fidelities in the interval ${[0.99,\,0.999]}$, with the central solid lines corresponding to a CZ gate fidelity of $0.995$.
}
\label{fig:oqft}
\end{figure*}

Here, we consider the construction where the OQFT is built exclusively from exact $\mathrm{QFT}_m^\dagger$ and $\mathrm{QFT}_{2m}$ subcircuits~\cite{Kahanamokumeyer2025}. 
Schematically, the OQFT on ${n=K\times m}$ qubits is composed of an initial and final layer of $\mathrm{QFT}_{2m}$ blocks, which are offset by $m$ qubits relative to each other, and an intermediate layer of $\mathrm{QFT}_m^\dagger$ blocks [see Fig.~\ref{fig:oqft}(a)]:
\begin{equation}
\begin{aligned}
\mathrm{OQFT}_{n}
&=
\bigotimes_{j=0}^{\left\lfloor K/2\right\rfloor-1}
\mathrm{QFT}_{2m}^{(2j+1,\,2j+2)} \cdot
\bigotimes_{j=1}^{K-2}
\mathrm{QFT}_{m}^{\dagger\,(j)}
\\
&\quad\cdot
\bigotimes_{j=1}^{\left\lceil K/2\right\rceil-1}
\mathrm{QFT}_{2m}^{(2j-1,\,2j)} \, .
\end{aligned}
\end{equation}
Here, the superscripts label the $m$-qubit blocks on which each QFT acts, with the blocks indexed from $1$ to $K$, while $\otimes$ indicates that the corresponding blocks within each layer are executed in parallel. 
% The floor and ceiling functions, $\lfloor\cdot\rfloor$ and $\lceil\cdot\rceil$, account for the parity of $K$ and ensure that the two $\mathrm{QFT}_{2m}$ layers cover the appropriate pairs of blocks. 
This blockwise construction allows large parts of the circuit to run in parallel, substantially reducing circuit depth. 
In particular, choosing $m=\mathcal{O}(\log(n/\epsilon))$ gives a logarithmic depth of $\mathcal{O}(\log(n/\epsilon))$, while the gate count remains of order $\mathcal{O}(n\log(n/\epsilon))$.

Since the Parity Twine framework provides a highly efficient implementation of arbitrary exact QFT circuits [cf.~Sec.~\ref{sec:qft}], each constituent $\mathrm{QFT}_{m}^\dagger$ and $\mathrm{QFT}_{2m}$ block can be directly implemented through its Parity Twine realization without modifying the overall OQFT structure.
One subtlety is that efficient QFT implementations, including the PTN realization, generally implement the transform up to a permutation of the qubits [cf.~Fig.~\ref{fig:ptn_circuits}], which has to be handled.

Building on the PTN-based exact QFT construction presented in Sec.~\ref{sec:qft}, we now co-design a highly efficient OQFT implementation for NAQCs.
Although both the linear and ladder QFT layouts serve as excellent building blocks, we specifically focus on the ladder configuration, which provides the most favorable performance among the considered layouts for the exact QFT.
We consider even $m$ and use rectangular patches of ${m = (m/2) \times 2}$ atoms.
We arrange the $m$-patches into a layout of 2 columns and $K/2$ rows, 
setting the inter-patch spacing to be the same as the intra-patch spacing.
This leads to a rectangular grid of $K/2 \times 2m$ atoms, as shown in Fig.~\ref{fig:oqft}(a), where the $m$-patches are distinguished by colors.
With this arrangement, each $\mathrm{QFT}_m^\dagger$ operation can be performed within a single $m$-patch using a horizontal Twine ladder.
Similarly, the $\mathrm{QFT}_{2m}$ operations can be efficiently performed on two horizontally neighboring $m$-patches by combining them into a single PTN ladder [see Fig.~\ref{fig:oqft}(a)]. 

The ladder PTN QFT reverses the qubit ordering at the end of the circuit [cf.~Fig.~\ref{fig:ptn_circuits}]. After the first layer of ${\mathrm{QFT}_{2m}}$ blocks, the qubit ordering within the ladders has consequently been reversed, as shown in Fig.~\ref{fig:oqft}(a) at time step $t_1$. Importantly, this reversal does not mix the qubit labels between individual $m$-patches, but only exchanges the left and right columns of $m$-patches and reverses the patch-intrinsic qubit ordering [see colored blocks in Fig.~\ref{fig:oqft}(a)]. The subsequent layer of $\mathrm{QFT}_m^\dagger$ operations can thus be performed directly, without changing the qubit arrangement or swapping qubit information. This intermediate layer restores the original qubit ordering within the $m$-patches, since the $\mathrm{QFT}_m^\dagger$ operations again reverse the ordering. At this point, the two columns of $m$-patches have effectively been swapped, such that exchanges of nearest-neighbor $m$-patch are sufficient to obtain the required ordering for the final layer of ${\mathrm{QFT}_{2m}}$ operations\footnote{Note that there can be $m$-patches on which no intermediate ${\mathrm{QFT}_m^\dagger}$ is applied. In these patches, the original qubit ordering is therefore not restored. However, this is not a problem, because these patches do not participate in the final layer of ${\mathrm{QFT}_{2m}}$ operations.}.

Crucially, the required patch exchanges can be implemented by collectively translating one of the two columns of patches, as shown at the bottom of Fig.~\ref{fig:oqft}(a) at time step $t_3$. This operation is naturally supported in both mobile and quasi-static neutral-atom architectures, where entire clusters of atoms can be translated in parallel with minimal overhead~\cite{Bluvstein2024, Rines2026}. After the final layer, the original qubit ordering can be restored through nearest-neighbor patch exchanges, which can again be implemented through collective transport of one of the two columns of patches, as shown at time step $t_5$ in the bottom of Fig.~\ref{fig:oqft}(a). The proposed collective transport achieves the target qubit ordering up to a left-right reversal, which must be accounted for in subsequent operations. If preserving the original left-to-right ordering is required, the collective transport can instead be implemented through column exchanges of atoms. The final patch exchanges can be omitted if restoring the original qubit ordering is not required by the target application.

Having established a highly efficient, tightly co-designed OQFT implementation for NAQCs, Fig.~\ref{fig:oqft}(b) summarizes the estimated fidelity of this implementation as a function of the total number of qubits, $n$, for different block sizes, $m$. 
Here, we use the CZSWAP-based (mobile) implementation of the ladder PTN; a benchmark against a quasi-static CZ-only implementation, where block movements are still performed via shuttling, is provided in App.~\ref{app:oqft_static_vs_mobile}. 
The fidelity estimates are obtained using the noise model and device parameters listed in Table~\ref{tab:params}.
For comparison, we also add a reference line for the fidelity of exact $\mathrm{QFT}_n$.
For small qubit numbers $n$, the exact QFT achieves the highest fidelity, as the overhead associated with partitioning the circuit into smaller blocks outweighs the reduction of circuit gate counts and depth. 
As $n$ increases, however, a clear crossover emerges beyond which the OQFT consistently outperforms exact QFT. 
The different scalings observed in Fig.~\ref{fig:oqft}(b) directly reflect the asymptotic resource requirements of the two algorithms. 
Since the dominant contribution to the circuit fidelity scales approximately as
${f \propto f_{\mathrm{2Q}}^{N_{\mathrm{2Q}}}}$, where $N_{\mathrm{2Q}}$ denotes the total number of two-qubit gates, shifting from quadratic (exact QFT) to nearly linear (OQFT) gate scaling, i.e. $\mathcal{O}(n\log(n/\epsilon))$, substantially slows the decay of overall fidelity with $n$.
The additional parallelization provided by the OQFT further suppresses depth-dependent error mechanisms, giving rise to the increasingly pronounced fidelity advantage observed for large $n$.

The improvement becomes even more significant for smaller block sizes $m$, reflecting the lower gate count and reduced depth of the constituent QFT blocks. This gain in implementation fidelity, however, comes at the expense of a smaller approximation subspace: decreasing $m$ reduces the size of the exact QFT blocks and consequently the class of states for which the OQFT faithfully reproduces the exact transformation~\cite{Kahanamokumeyer2025}. 
The block size $m$ therefore provides a tunable trade-off between implementation fidelity and the size of the subspace on which the OQFT remains exact.

An alternative formulation of the OQFT entirely avoids the use of ${\mathrm{QFT}_{2m}}$ blocks by utilizing only ${\mathrm{QFT}_{m}}$ blocks and inter-block controlled-phase rotations, offering further reductions in gate count and circuit depth~\cite{Kahanamokumeyer2025}. 
The Parity Twine framework provides an ideal platform for realizing this optimized variant, which is the subject of an upcoming work~\cite{parity_oqft_in_prep}.

\section{Summary and Outlook}
\label{sec:outlook}
In this work, we have presented a hardware-algorithm co-design strategy for neutral-atom quantum computing based on Parity Twine Networks. By adapting the Parity Twine framework to native neutral-atom gate sets and shuttling primitives, we developed efficient realizations based on both CZ and CZSWAP operations and analyzed their performance across different hardware layouts. Our results for the QFT algorithm show that PTNs substantially reduce entangling-gate count, atom movement, and transfer overhead while naturally supporting both static and shuttling-assisted architectures. 
This flexibility enables hardware-aware optimizations across gate count, circuit depth, and atom transport, leading to substantial improvements in the estimated fidelity of QFT implementations under diverse realistic hardware assumptions.

While we have focused on monolithic devices in this work, the PTN framework can also be extended to zone-based architectures~\cite{Bluvstein2022, Bluvstein2024, Reichardt2025}. 
Specifically, CZSWAP-based PTN implementations offer a highly shuttling-efficient method for realizing dense interaction graphs on these platforms. 
Since qubits must return to the storage zone after an entangling gate regardless, routing them to swapped rather than original positions incurs negligible additional overhead.

Looking beyond near-term algorithms, the drastic reduction in physical shuttling overhead obtained from PTNs has profound implications for fault-tolerant architectures based on the transport of QEC code patches~\cite{Bluvstein2024, Rines2026, Bluvstein2026, Ismail2026, Bhardwaj2026}. 
Specifically, the PTN approach will directly reduce the substantial transport overhead required for transversal logical entangling operations for algorithms implementing dense interaction graphs.

%The Parity Twine framework is particularly efficient for quantum algorithms with dense interaction graphs, where conventional hardware-aware compilation typically incurs significant routing overhead. 
Beyond the QFT and OQFT studied in this work, another prominent application is quantum optimization. In particular for QAOA, where cost Hamiltonians often require implementing a high density of two-qubit interactions~\cite{Koch2025}, the use of PTNs can substantially reduce gate count and circuit depth~\cite{Dreier2025}, boosting optimization performance~\cite{Montanezbarrera2025}.
The Parity Twine framework can also be extended to efficiently implement interaction graphs
with varying degrees of sparsity, as demonstrated in an upcoming publication~\cite{parity_oqft_in_prep}.

Another promising direction is Fourier-space quantum arithmetic, where the QFT serves as a central building block for operations such as addition and multiplication~\cite{Draper2000, Ruiz-Perez2017}. Since these arithmetic circuits contain QFTs as their dominant components, the substantial reductions in gate resources achieved by the PTN construction can translate directly into corresponding improvements for Fourier-based arithmetic. Moreover, the OQFT framework provides a natural route toward approximate arithmetic with further reductions in circuit depth. We briefly outline these opportunities in App.~\ref{app:arithmetic}.

PTNs are also a natural compilation framework for Instantaneous Quantum Polynomial-time (IQP) circuits, which consist of commuting gates that are diagonal in the ${\mathrm{X}}$ basis and can be expressed as dense parity polynomials. 
Despite their simple circuit structure, IQP circuits are believed to be classically hard to sample, making them attractive candidates for demonstrating quantum advantage~\cite{Marshall2024, Bremner2016}. At the same time, they can be efficiently trained using classical optimization techniques, which has recently motivated their use as expressive quantum generative models, see, e.g., Ref.~\cite{Recioarmengol2026, Banks2026, Tuysuz2026}. These examples illustrate that the Parity Twine framework provides a general hardware-aware compilation strategy for an important class of dense quantum circuits, while remaining adaptable to different native gate sets and hardware connectivities.

\begin{acknowledgements}
The authors gratefully acknowledge their colleagues at ParityQC for their technical contributions and software development efforts that supported this work. We also thank Andrew Byun for his helpful comments on the manuscript.

This research was funded in whole, or in part, by the Austrian Science Fund (FWF) SFB BeyondC Project No. F7108-N38 (DOI: 10.55776/F71). This project was supported by FFG Fundings (Project Nos. FO99918691 and FO999933929) as part of the international Eureka cooperation. This study was supported by the Austrian Research Promotion Agency (FFG Project No. FO999937388, FFG Basisprogramm).
This research is funded by the German Federal Ministry of Research, Technology and Space (BMFTR) within the project MUNIQC-ATOMS (Project No. 13N16080).
This publication has received funding under Horizon Europe programme HORIZON-CL4-2022-QUANTUM-02-SGA via the project 101113690 (PASQuanS2.1).
\end{acknowledgements}

\bibliography{references}

\appendix
\onecolumngrid

\section{CZ versus CZSWAP in mobile architectures}
\label{app:czswap}

\begin{figure}[!t]
\centering
\includegraphics[width=0.95\textwidth]{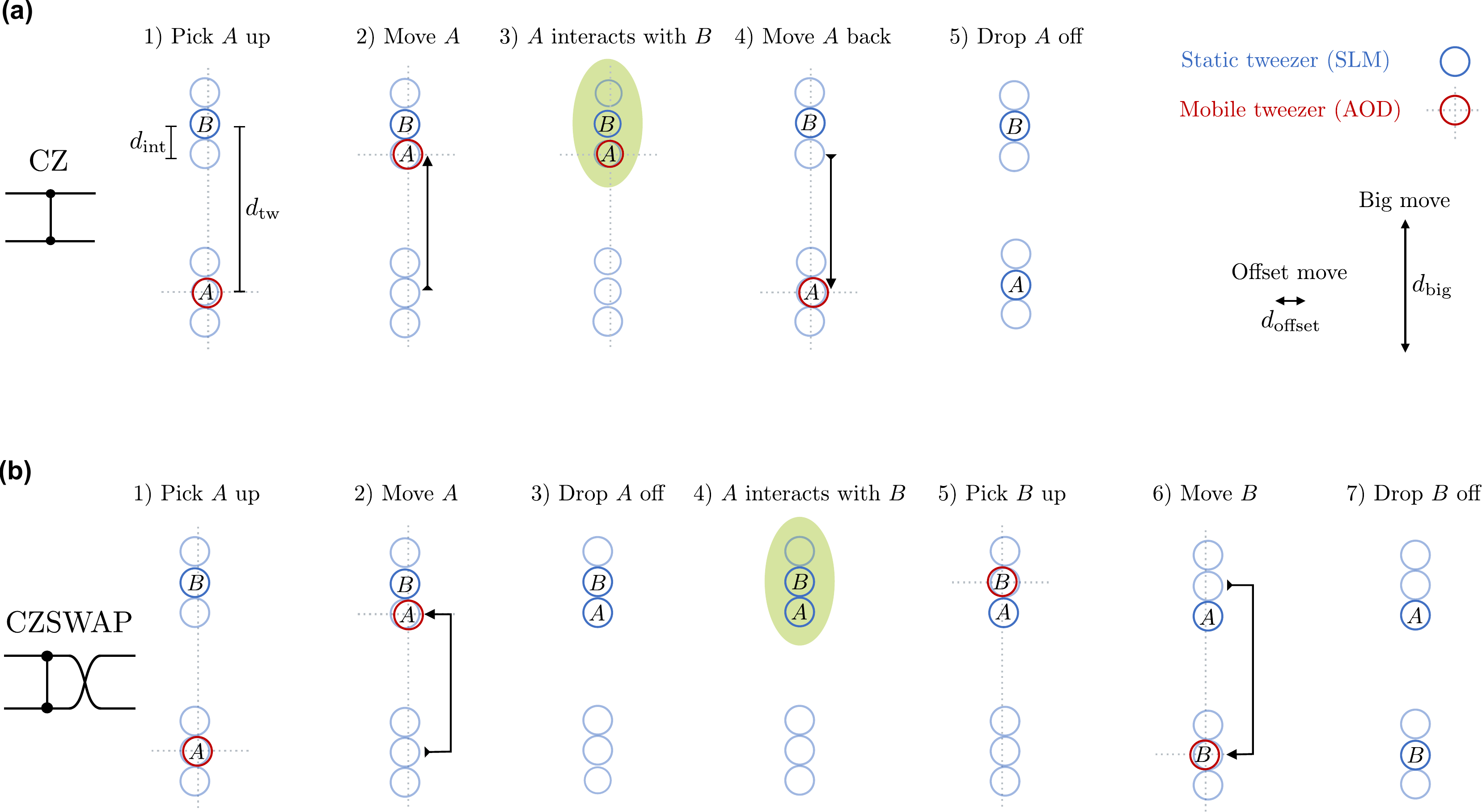}
\caption{Illustration of the implementation of (a) a Rydberg-mediated CZ gate and (b) a CZSWAP operation in mobile architectures. In both cases, atoms $A$ and $B$ are brought from their initial separation $d_{\mathrm{tw}}$ to the CZ interaction distance $d_{\mathrm{CZ}}$. For the CZ, atom $A$ is subsequently returned to its original position. For the CZSWAP, the transport instead exchanges the positions of $A$ and $B$, with additional short offset moves used to avoid collisions and ensure available SLM traps for atom transfer.}
\label{fig:czswap}
\end{figure}

Figure~\ref{fig:czswap} illustrates the physical implementation of a
Rydberg-mediated CZ gate and a CZSWAP operation in the mobile architecture
considered in this work. In both cases, two atoms $A$ and $B$, initially
separated by $d_{\mathrm{tw}}$, are brought to the interaction distance
$d_{\mathrm{int}}$ using an AOD tweezer. We assume that CZ gates can be
performed directly between atoms held in the AOD and SLM traps, see step 3
in Figure~\ref{fig:czswap}(a). For a CZ gate, atom $A$ is therefore
transferred to the AOD, transported into proximity with $B$, and, after the
interaction, transported back and transferred to the SLM array.

For a CZSWAP, the return transport of atom $A$ is instead used to exchange
the positions of the two atoms. After the CZ interaction, atom $A$ is
dropped off at an available SLM trap while atom $B$ is captured by the AOD
and transported to the original position of $A$, as depicted in
Figure~\ref{fig:czswap}(b). This realizes the desired CZ followed by a
SWAP without requiring a separate gate-based SWAP operation. The exchange
requires additional short offset moves. In our model, we assume two
offset moves per big move, resulting in four offset moves per CZSWAP.
These offset moves account for the need to maintain empty SLM traps for
dropping off atoms $A$ and $B$ and may require horizontal displacements
to avoid collisions between atoms during the exchange.

Here, a \emph{big move} denotes the transport distance ${d_{\mathrm{big}} = d_{\mathrm{tw}}-d_{\mathrm{int}}}$, while an \emph{offset move} denotes the additional short displacement
required during the exchange. The total transport distances associated
with a CZ and CZSWAP operation are therefore
\begin{equation*}
    \begin{split}
    d_{\mathrm{CZ}} &= 2d_{\mathrm{big}},\\
    d_{\mathrm{CZSWAP}} &= 2d_{\mathrm{big}}+4d_{\mathrm{offset}}.
    \end{split}
\end{equation*}
For the parameters used in this work, summarized in
Table~\ref{tab:params} in the main text, we set $d_{\mathrm{int}}=3\,\mu\mathrm{m}$ to
enable fast, high-fidelity CZ gates while suppressing crosstalk at the
neighboring tweezer separation $d_{\mathrm{tw}}=15\,\mu\mathrm{m}$, and
$d_{\mathrm{offset}}=2\,\mu\mathrm{m}$ to ensure sufficient separation
from neighboring traps, given the spatial resolution of typical tweezer arrays. This gives
$d_{\mathrm{CZ}}=24\,\mu\mathrm{m}$ and
$d_{\mathrm{CZSWAP}}=32\,\mu\mathrm{m}$.

Thus, compared with two independently executed CZ gates, a CZSWAP replaces
one CZ gate and its associated transport by the exchange of the atomic
positions. For decomposition~(i) [see Fig.~\ref{fig:czswap}(b) in the main text], the two CZ gates require four transfer
operations and four big moves. The number of transfer operations can,
however, be reduced to two by skipping the intermediate drop-off and
pick-up steps between consecutive CZ gates. In contrast, the CZSWAP
requires additional offset moves but only a single entangling CZ operation.
The CZSWAP therefore trades additional transport overhead for a reduction
in the number of CZ gates and associated Rydberg stages. This trade-off is
particularly favorable when CZ errors and Rydberg crosstalk are relatively
strong. Conversely, when CZ gates have sufficiently high fidelity and
crosstalk is negligible, the additional transport and possible transfer
overhead associated with CZSWAP can outweigh this advantage.

\section{Resource Scaling of PTN-Based QFT}
\label{app:resource_scaling}

Table~\ref{tab:ptn_scaling} summarizes the gate-count and circuit-depth scaling of the PTN-based QFT for the linear, ladder, and square layouts. Here, GR denotes global single-qubit rotations, RZ denotes local $R_Z$ rotations, and the total numbers of single- and two-qubit gates are defined as
\begin{equation*}
\begin{split}
N_{\mathrm{1Q}}&=N_{\mathrm{GR}}+N_{\mathrm{RZ}},\\
N_{\mathrm{2Q}}&=N_{\mathrm{CZ}}+N_{\mathrm{CZSWAP}}.
\end{split}
\end{equation*}
For all three layouts, the total two-qubit gate count scales quadratically with the number of qubits, while the single-qubit gate count scales linearly. Despite this quadratic gate-count scaling, the circuit depth remains linear in $n$ for all considered layouts. The different layouts primarily redistribute the two-qubit gates between CZ and CZSWAP operations and modify the associated depth and single-qubit overhead.

Table~\ref{tab:ptn_scaling} shows that the PTN construction achieves the same asymptotic two-qubit gate count, 
\begin{equation*}
N_{\mathrm{2Q}}=\frac{1}{2}n^2+\frac{1}{2}n-1,
\end{equation*}
for all three layouts, while redistributing the gates between CZ and CZSWAP operations according to the available connectivity. The linear layout requires the largest number of CZSWAP gates, whereas increased connectivity progressively replaces CZSWAP operations with CZ gates. This redistribution, however, does not reduce the overall two-qubit gate depth for the linear and ladder layouts, which remains $2n-1$. The ladder therefore provides a particularly favorable compromise, achieving the same two-qubit gate count and depth as the linear layout while substantially reducing the number of CZSWAP operations. In contrast, the square layout further reduces the CZSWAP count at the expense of increased single-qubit and two-qubit depth. Thus, increasing connectivity does not necessarily improve all circuit resources simultaneously; rather, it enables a trade-off between gate type, depth, and the associated hardware-dependent implementation costs.

\begin{table}[!b]
\centering
\setlength{\tabcolsep}{8pt}
\renewcommand{\arraystretch}{1.25}

\caption{Scaling of gate counts and gate depths for PTN-based QFT circuits for the linear, ladder, and square layouts. Here, $n$ denotes the number of qubits. GR denotes global $XY$ rotations, RZ denotes local $R_Z$ rotations, and 1Q and 2Q denote all single- and all two-qubit gates, respectively. The expressions are obtained by polynomial fitting to numerical data for $20\leq n\leq100$, considering only complete square grids with $n=l^2$ for the square layout. The fits perfectly reproduce the data except for the CZ and CZSWAP gate counts for the square layout, which are approximate fits.}

\label{tab:ptn_scaling}

\begin{tabular}{ll|ccc|ccc}
\toprule
\textbf{Layout} &
\textbf{Metric} &
$\bm{\mathrm{GR}}$ &
$\bm{\mathrm{RZ}}$ &
$\bm{\mathrm{1Q}}$ &
$\bm{\mathrm{CZ}}$ &
$\bm{\mathrm{CZSWAP}}$ &
$\bm{\mathrm{2Q}}$ \\
\midrule

\multirow{2}{*}{Linear}
& Count
& $4n-3$
& $23n-194$
& $27n-197$
& $n-1$
& $\frac{1}{2}n^2-\frac{1}{2}n$
& $\frac{1}{2}n^2+\frac{1}{2}n-1$
\\
& Depth
& $4n-3$
& $2n$
& $6n-3$
& $2$
& $2n-3$
& $2n-1$
\\[8pt]
\hline
\multirow{2}{*}{Ladder}
& Count
& $4n-3$
& $23n-194$
& $27n-197$
& $\frac{1}{4}n^2+\frac{1}{2}n-1$
& $\frac{1}{4}n^2$
& $\frac{1}{2}n^2+\frac{1}{2}n-1$
\\
& Depth
& $4n-3$
& $2n$
& $6n-3$
& $\frac{1}{2}n+1$
& $\frac{3}{2}n-2$
& $2n-1$
\\[8pt]
\hline
\multirow{2}{*}{Square}
& Count
& $8n-33$
& $23n-194$
& $31n-227$
& $\approx\frac{1}{4}n^2+2.8n-35$
& $\approx\frac{1}{4}n^2-2.3n+34$
& $\frac{1}{2}n^2+\frac{1}{2}n-1$
\\
& Depth
& $8n-33$
& $4n-15$
& $12n-48$
& $13$
& $4n-25$
& $4n-12$
\\

\bottomrule
\end{tabular}
\end{table}

\section{CZ-Fidelity Threshold for Static and Mobile OQFT}
\label{app:oqft_static_vs_mobile}

\begin{figure}[!t]
\centering
\includegraphics[width=0.6\textwidth]{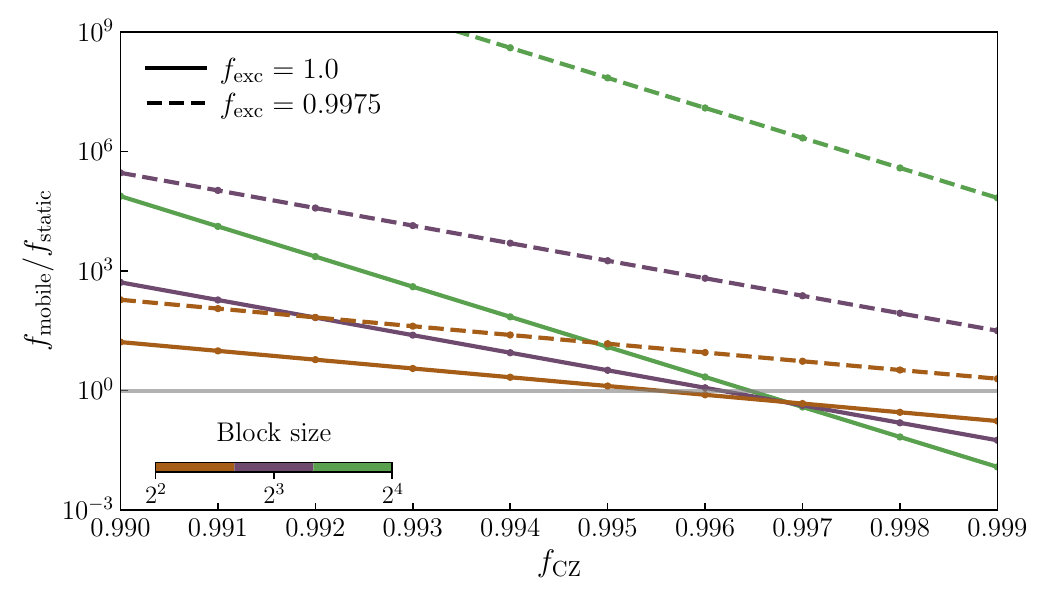}
\caption{Ratio of the estimated fidelities of the mobile and static PTN-based $\mathrm{OQFT}_{64}$ circuits as a function of the CZ-gate fidelity for different block sizes $m$. The curves show the ratio in the presence and absence of a $0.25\%$ Rydberg excitation error. The gray horizontal line indicates a fidelity ratio of one, separating the regimes in which the mobile or static architecture is preferred. In the presence of crosstalk, the mobile implementation remains advantageous even for $f_{\mathrm{CZ}}=0.999$, whereas in the crosstalk-free case, a characteristic CZ-fidelity threshold emerges at approximately ${f_{\mathrm{CZ}}\approx0.997}$, above which the static implementation is advantageous for all considered values of $m$.}
\label{fig:oqft_fid_ratio}
\end{figure}

As a complementary analysis to the OQFT results presented in Sec.~\ref{sec:oqft}, we further investigate the trade-off between atom transport and entangling-gate errors by comparing the estimated fidelities of the mobile and static PTN-based implementations as a function of the CZ-gate fidelity. Figure~\ref{fig:oqft_fid_ratio} shows the ratio of the mobile to static fidelity for $\mathrm{OQFT}_{64}$ for several block sizes $m$, considering both the presence and absence of Rydberg crosstalk. A ratio larger than one indicates that the mobile implementation is more favorable, whereas a ratio below one indicates that the static implementation achieves higher fidelity.

In the presence of Rydberg crosstalk, the mobile implementation consistently outperforms the static implementation over the considered range of CZ-gate fidelities up to $f_{\mathrm{CZ}}=0.999$. This advantage results from the reduced number of entangling operations in the mobile realization: although atom transport introduces additional transfer and shuttling errors, the reduction in CZ gates and Rydberg-excitation stages provides a net fidelity benefit when crosstalk errors are present.
In the absence of crosstalk, the competition between the two architectures gives rise to a characteristic CZ-fidelity threshold. Above approximately ${f_{\mathrm{CZ}}\approx0.997}$, the static implementation can outperform the mobile one independently of the block size $m$.

\section{Quantum Arithmetic}
\label{app:arithmetic}

The PTN-based QFT can also serve as a building block for Fourier-space
quantum arithmetic~\cite{Draper2000, Ruiz-Perez2017}. In general, Fourier-space operators can be expressed in
the form
\begin{equation*}
A = \mathrm{QFT}^{\dagger} D\,\mathrm{QFT},
\end{equation*}
where $D$ is a diagonal operator in the Fourier basis. For quantum arithmetic, $D$ is a diagonal unitary whose phases encode the corresponding arithmetic function. Similar Fourier-space decompositions have also been used beyond
arithmetic, for example in quantum algorithms for solving partial
differential equations, where the QFT can be used to diagonalize the
relevant differential operators~\cite{Wright2024, Lubasch2025}.

As a simple example, we consider the Draper adder~\cite{Draper2000},
which performs addition by transforming the input register to the Fourier
basis, applying a phase gradient $D_a$ corresponding to the classical integer
$a$ to be added, and transforming back. The circuit takes the form
\begin{equation*}
\begin{split}
\mathrm{ADD}_a &= \mathrm{QFT}^{\dagger}D_a\mathrm{QFT}, \\
\rm{with}\quad D_a|b\rangle &= \exp\left( 2\pi iab/2^n \right)|b\rangle,
\end{split}
\end{equation*}
implementing $|b\rangle \mapsto |a+b\mod{2^n}\rangle$. Here, $D_a$ consists
only of single-qubit $Z$ rotations whose angles are determined by the binary
representation of $a$. Both QFT transformations can be implemented using the PTN construction, while the intermediate phase rotations are applied directly to
the corresponding logical qubits. Since the two QFTs dominate the resource
requirements, the resulting PTN-based Draper adder requires twice the gate
count and circuit depth of a single PTN-based QFT, with only up to $n$
single-qubit phase rotations.

The same Fourier-space construction also enables addition of two quantum
registers~\cite{Draper2000, Ruiz-Perez2017}. Let the two $n$-qubit registers encode the integers $a$ and $b$,
respectively. The desired quantum-quantum adder can then be written as
\begin{equation*}
\begin{split}
\mathrm{ADD}=&(\mathbb{I}\otimes\mathrm{QFT}^{\dagger})
D_{ab}(\mathbb{I}\otimes\mathrm{QFT}),\\
\rm{with}\quad &
D_{ab}|a\rangle|b\rangle
=\exp\left( 2\pi iab/2^{2n}\right) |a\rangle|b\rangle,
\end{split}
\end{equation*}
implementing $|a\rangle|b\rangle \mapsto |a\rangle|a+b\mod{2^n}\rangle$.
In contrast to the classical-quantum case, where $D_a$ consists only of
single-qubit phase rotations, $D_{ab}$ requires
$\mathcal{O}(n^2)$ two-qubit phase interactions between the two registers.
While the number of interactions remains quadratic, the structured
phase-interaction graph can be implemented with a reduced gate-count
prefactor compared with a generic all-to-all realization. Importantly, the
PTN concept can be extended to efficiently implement phase-interaction graphs
with varying degrees of sparsity, as demonstrated for related
phase-interaction operators in an upcoming publication~\cite{parity_oqft_in_prep}.

Multiplication can be constructed analogously by encoding the corresponding
product in a diagonal phase operator implementing $\exp{(2\pi iabc/2^{2n}})$
where $a$ may represent either a classical integer or a quantum register,
with the product of $a$ and $b$ accumulated into the output register $c$.
More efficient QFT-based modular multiplication can be obtained by
incorporating classical multiplication techniques such as the Toom-Cook
algorithm, as described, for example, in Ref.~\cite{Kahanamokumeyer2024}.
The optimistic-QFT framework provides a natural route to approximate these
circuits: replacing the QFTs appearing in the QFT-based multiplier by OQFTs
yields an optimistic multiplier with reduced depth while retaining the
low-ancilla structure of the underlying construction. Such optimistic
multipliers have been shown to be sufficient for quantum algorithms such as
Shor's factoring algorithm, illustrating that the optimistic approximation
can be propagated through nontrivial arithmetic computations~\cite{Kahanamokumeyer2025}.

These observations suggest a broader application of the Parity Twine framework to
Fourier-based arithmetic, complementing previous work on quantum arithmetic
with parity-based operations~\cite{Fellner2022b}. In particular, the PTN
implementation of the QFT can be combined with co-designed OQFT
constructions to realize optimistic adders and multipliers while retaining
the hardware-aware advantages of PTNs. A detailed resource analysis of
PTN-based optimistic arithmetic, including modular addition and
multiplication, is left for future work.

\end{document}